\documentclass{aa}  
\usepackage{graphicx,natbib}
\usepackage{txfonts}
\usepackage{color}
\newcommand{\Msun}{\ensuremath{\mathit{M}_{\odot}}}
\newcommand{\Ni}{\ensuremath{^{56}\mathrm{Ni}}}
\newcommand{\Co}{\ensuremath{^{56}\mathrm{Co}}}

\newcommand{\kmps}{\mathrm{km~s^{-1}}}
\newcommand{\w}{\mathrm{wind}}
\newcommand{\csm}{\mathrm{csm}}
\newcommand{\ej}{\mathrm{ej}}

\begin{document} 

   \authorrunning{Moriya et al.}
   \titlerunning{Electron-capture SNe within super-AGB wind}

   \title{
Electron-capture supernovae exploding within \\
their progenitor wind
 }

   \subtitle{}

   \author{Takashi J. Moriya\inst{1},
Nozomu Tominaga\inst{2,3},
Norbert Langer\inst{1},
Ken'ichi Nomoto\inst{3,4}, \\
Sergei I. Blinnikov\inst{5,3},  \and
Elena I. Sorokina\inst{6}
          }

   \institute{
Argelander Institute for Astronomy, University of Bonn,
Auf dem H\"ugel 71, D-53121 Bonn, Germany
   \\
              \email{moriyatk@astro.uni-bonn.de} 
\and
Department of Physics, Faculty of Science and Engineering, Konan
University, 8-9-1 Okamoto, Kobe, Hyogo 658-8501, Japan
\and
Kavli Institute for the Physics and Mathematics of the Universe (WPI),
Todai Institutes for Advanced Study, The University of Tokyo,
5-1-5 Kashiwanoha, Kashiwa, Chiba 277-8583, Japan
\and
Hamamatsu Professor
\and
Institute for Theoretical and Experimental Physics,
Bolshaya Cheremushkinskaya 25, 117218 Moscow, Russia
\and
Sternberg Astronomical Institute, M.V. Lomonosov Moscow State
University,
Universitetski pr. 13, 119992 Moscow, Russia
             }

   \date{Received 23 May 2014 / accepted 17 July 2014}

\abstract{
The most massive stars on the asymptotic giant branch (AGB), so called
super-AGB stars, are thought to produce supernovae triggered by electron
captures in their degenerate O+Ne+Mg cores. Super-AGB stars are expected
to have slow winds with high mass-loss rates, so their circumstellar
density is high. The explosions of super-AGB stars are therefore
presumed to occur in this dense circumstellar environment. We provide
the first synthetic light curves for such events by exploding realistic
electron-capture supernova progenitors within their super-AGB winds. We
find that the early light curve, i.e. before the recombination wave
reaches the bottom of the hydrogen-rich envelope of supernova ejecta
(the plateau phase), is not affected by the dense wind. However, after
the luminosity drop following the plateau phase, the luminosity remains
much higher when the super-AGB wind is taken into account. We compare
our results to the historical light curve of SN\,1054, the progenitor of
the Crab Nebula, and show that the explosion of an electron-capture
supernova within an ordinary super-AGB wind can explain the observed
light curve features. We conclude that SN\,1054 could have been a
Type\,IIn supernova without any extra extreme mass loss which was
previously suggested to be necessary to account for its early high
luminosity. We also show that our light curves match Type\,IIn
supernovae with an early plateau phase (so-called `Type IIn-P') and
suggest that they are electron-capture supernovae within super-AGB
winds. Although some electron-capture supernovae can be bright in the
optical spectral range due to the large progenitor radius,
their X-ray luminosity from the interaction does not necessarily
get as bright as other Type IIn supernovae
whose optical luminosities are also powered by the interaction.
Thus, we suggest that
optically-bright X-ray-faint Type IIn supernovae can emerge from
electron-capture supernovae.
Optically-faint Type IIn supernovae, such as SN 2008S, can also originate from 
electron-capture supernovae if their hydrogen-rich envelope masses are small.
We argue that some of them can be observed as `Type IIn-b' supernovae due to
the small hydrogen-rich envelope mass.
}

 


   \maketitle
%

\section{Introduction}\label{introduction}
Massive stars are known to explode as supernovae (SNe) because of the
central core collapse. 
Stars with the zero-age
main-sequence mass above around 10 \Msun\ trigger the
core collapse after the formation of an Fe core.
However, it has been suggested that massive stars whose mass is slightly
smaller than the mass required to form the Fe core can still make
an electron degenerate O+Ne+Mg core which can trigger core collapse through
the electron-capture reactions
\citep[e.g.,][]{miyaji1980,nomoto1982,nomoto1984,nomoto1985,nomoto1987,miyaji1987}.
Core-collapse SNe triggered by the electron-capture reactions are called
electron-capture SNe (ecSNe).

The ecSN progenitors are super-asymptotic giant branch (super-AGB) stars
at the onset of the core collapse if they are single stars.
Since there are many uncertainties 
in thermal pulses, mass loss, etc.,
in the evolution of super-AGB stars,
the exact mass range of ecSN progenitors,
or even the existence of them,
is theoretically not well-constrained
\citep[e.g.,][]{langer2012,poelarends2008,siess2007,siess2010,jones2013,takahashi2013}.
It is also suggested that the binary evolution can help a star exploding
as an ecSN \citep{podsiadlowski2004,nomoto1985}.
While the exact mass range or the existence of ecSN progenitors remains
uncertain, several first-principle numerical simulations of
the O+Ne+Mg core collapse succeeded in predicting
the observational properties of ecSNe.
The ecSNe are predicted to have
small explosion energy ($\sim10^{50}$ erg) and small \Ni\ production
($\sim 10^{-3}$ \Msun) by the neutrino-driven explosion
\citep[][]{kitaura2006,burrows2007,janka2008,hoffman2008,wanajo2009}.

Several SNe have been related to ecSNe
based on the predicted ecSN properties
\citep[e.g..][]{botticella2009,thompson2009,pumo2009,kawabata2010}.
The most famous SN related to an ecSN is SN 1054 or the Crab Nebula
(see Section \ref{sec1054}).
However, while several numerical simulations of the explosions of ecSNe
have been performed by using the realistic progenitor model,
there is little theoretical study on the
LC properties of ecSNe. Recently,
\citet[][TBN13 hereafter]{tominaga2013} performed numerical LC
calculations by using realistic ecSN progenitor model and provided
the LC properties of ecSNe.
They have shown that ecSNe can be bright ($\sim -17$ mag in optical bands)
in spite of the small predicted explosion energy because of the
large progenitor radius and the small H-rich envelope mass.
Super-AGB stars are known to have high mass-loss rates
($\sim 10^{-4}$ $M_\odot~\mathrm{yr^{-1}}$, e.g., \citealt{poelarends2008})
and the ecSN progenitors can lose a large fraction of their H-rich envelope as
a result of the large mass loss during the super-AGB phase,
which is quite distinct from a-little-more massive stars
causing the Fe core collapse.

An important characteristics of ecSNe which TBN13 did not consider 
is the existence of the super-AGB wind around the ecSN progenitors.
The ecSN progenitors should be within the circumstellar medium (CSM)
created by the super-AGB wind when they explode
if they are single stars.
In addition to the high mass-loss rates,
the wind velocities of super-AGB stars are
$\sim 10\ \kmps$, making the CSM density as high as those expected
for Type IIn SNe (SNe IIn) in which we see the effect of the CSM interaction
in their observational
properties (e.g., \citealt{kiewe2012,taddia2013,fransson2013,moriya2014};
see Section \ref{seccsm}). It is even suggested that most of SNe IIn may come
from the low mass stars which are consistent with ecSNe
\citep{habergham2014,anderson2012}.

The fact that ecSNe can be bright despite the small explosion energy
was already suggested by \citet{vanveelen2010} by modeling the interaction
between ecSN ejecta and expected dense CSM.
\citet{smith2013} also discussed the effect of dense CSM on ecSN LCs.
He suggested that the early large luminosity of SN 1054 from an ecSN indicates
the existence of confined dense CSM from eruptive mass loss
(see Section \ref{sec1054} for the discussion of SN 1054).
However, as is shown by TBN13; \citet{vanveelen2010},
ecSNe can be as bright as SN 1054 in spite of the low explosion energy
because of the expected large progenitor radius and small mass
of the H-rich envelope.

The previous numerical simulations of ecSN LCs do not take either the
realistic ecSN progenitor model \citep{vanveelen2010} or the existence
of the dense CSM around the ecSN progenitors (TBN13) into account.
In this paper, we combine the two previous LC studies of ecSNe and
perform the numerical LC simulation of ecSNe exploding in the super-AGB wind.
We provide realistic ecSN LCs to compare with observations.

The rest of this paper is organized as follows.
At first, in Section \ref{secprogenitor}, we explain the progenitors
and their CSM we use for our LC modeling of ecSNe as well as
the properties of the SN explosions.
Then, we show
the synthesized LCs in Section \ref{secLC}. The LCs are compared to
observed SN LCs with special focus on SN 1054 and so-called `Type IIn-P'
SNe in Section \ref{secobs}. We conclude this paper in Section \ref{secconclusion}.

\section{Initial conditions}
\subsection{Progenitors and circumstellar media}\label{secprogenitor}
\subsubsection{Electron-capture SN progenitors}
The super-AGB ecSN progenitors we use in this work is the same as
those presented in TBN13.
The progenitor models are constructed by attaching
a H-rich envelope \citep{nomoto1972} to the thin He layer on
the 1.377 $M_\odot$ O+Ne+Mg ecSN progenitor
obtained by \citet{nomoto1982,nomoto1984,nomoto1987}.
TBN13 constructed several ecSN progenitors with different envelope
masses ($M_\mathrm{env}$) and hydrogen fractions ($X_\mathrm{env}$).
In this work, we take the model with $M_\mathrm{env}=3.0$ $M_\odot$ and
$X_\mathrm{env}=0.2$ as a fiducial model.
The LC from the model was compare to that of SN 1054 in TBN13.
The hydrogen fraction is rather low in the model.
The hydrogen fraction affects the early plateau length in LCs, but
the differences between the plateau lengths of
the $X_\mathrm{env}=0.2$ models and 
the $X_\mathrm{env}=0.7$ models are less than 20 days (TBN13).
The plateau luminosity is not strongly affected by the composition.
The radius and luminosity of the model are $7\times 10^{13}$ cm
($10^3$ $R_\odot$) and $3\times 10^{38}$ $\mathrm{erg~s^{-1}}$
($8\times 10^4$ $L_\odot$), respectively. See TBN13 for the detailed
properties of the progenitors.

\subsubsection{Circumstellar media}\label{seccsm}
We put the CSM expected from super-AGB winds outside the aforementioned ecSN
progenitors.
However, the mass loss from super-AGB stars is uncertain.
\citet{poelarends2008} adopted several mass-loss prescriptions to follow
the super-AGB evolution and found that the differences in the mass-loss
prescriptions do not significantly affect the final results.
The mass-loss rates are around $10^{-4}$ $M_\odot~\mathrm{yr^{-1}}$
\citep[][]{poelarends2008}. We adopt three mass-loss rates in our LC
calculations, i.e., $10^{-4}$, $5\times 10^{-5}$, and $10^{-5}$
$M_\odot~\mathrm{yr^{-1}}$. Although we construct the CSM assuming that
the mass-loss rates are constant, super-AGB stars 
experience thermal pulses and the mass-loss rates can alter
significantly in a short timescale \citep[e.g.,][]{jones2013,lau2012}.

Another important quantity we need to know to estimate the CSM density
is the wind velocity $v_\w$. The escape velocity of our fiducial
progenitor is 41 $\mathrm{km~s^{-1}}$.
The actual terminal wind velocity is, again, uncertain.
\citet{vanveelen2010} argued that the wind velocity can be as low as 30\%
of the escape velocity based on the formulation of \citet{eldridge2006}.
Here, we take $v_\w=20$ $\mathrm{km~s^{-1}}$ as a representative wind
velocity as was assumed in \citet{vanveelen2010}.

Given the mass-loss rate ($\dot{M}$) and wind velocity ($v_\w$),
the CSM density $\rho_\csm(r)$ is set as
\begin{equation}
 \rho_\csm (r)=\frac{\dot{M}}{4\pi v_\w}r^{-2},\label{winddensity}
\end{equation}
where $r$ is radius.
The super-AGB phase
is assumed to continue for $\sim 10^4$ years \citep[e.g.,][]{poelarends2008} and
the dense CSM should reach $\sim 6\times 10^{17}$ cm in the super-AGB phase.
However, the CSM radius boundary is arbitrarily set at $2\times 10^{16}$ cm in the
initial condition of the numerical simulations we performed.
Thus, the effect of the dense CSM on the LCs is expected to 
last longer than we obtain, as will be discussed in the following sections.

It is an important feature that the CSM density of the super-AGB wind
is similar to those estimated in the CSM of typical SNe IIn.
The CSM density is proportional to $\dot{M}/v_\w$
(Eq. \ref{winddensity}). 
The typical CSM properties of SNe IIn are estimated to be 
$\dot{M}\sim 10^{-3}$ $M_\odot~\mathrm{yr^{-1}}$ with $v_\w\sim 100$
$\mathrm{km~s^{-1}}$ \citep[e.g.,][]{kiewe2012,taddia2013,fransson2013,moriya2014}.
While the mass-loss rates of the super-AGB stars
($\sim 10^{-4}$ $M_\odot~\mathrm{yr^{-1}}$) are about one order of
magnitude smaller than those estimated for SNe IIn, the wind velocity
$\sim 10$ $\mathrm{km~s^{-1}}$ is also one order of magnitude smaller.
Thus, the CSM density from the super-AGB wind is presumed to be high
enough to affect the LCs and spectra of ecSNe as much as we observe in
SNe IIn.

\subsection{Explosion properties and light curve calculations}
The explosions of super-AGB stars are followed in the same way as in TBN13.
The explosions of the ecSN progenitors are numerically followed with
a one-dimensional multigroup radiation hydrodynamics code
\texttt{STELLA} \citep[e.g.,][]{blinnikov2006,blinnikov1998}.
The explosions are initiated by putting thermal energy at the center.
The explosion energy is set as $1.5\times 10^{50}$
erg, which is based on the numerical simulations performed with the first
principles \citep{kitaura2006}, unless otherwise mentioned. \texttt{STELLA} does not follow
the explosive nucleosynthesis and we need to provide the chemical yields
from the explosive nucleosynthesis to obtain the opacity and
the energy input from the nuclear decay of \Ni.
We use the chemical yields from a nucleosynthesis calculation of the
self-consistent ecSN explosion performed with the same O+Ne+Mg core
(Model ST in \citealt{wanajo2009}). $2.5\times 10^{-3}$ $M_\odot$ of
\Ni\ is produced in the model.
We assume the envelope is not affected by the explosive nucleosynthesis.
Finally, TBN13 investigated the effect of the central pulsar to the LC but we do
not take it into account in this paper.

\section{Light curve properties}\label{secLC}
\begin{figure}
\centering
\includegraphics[width=\columnwidth]{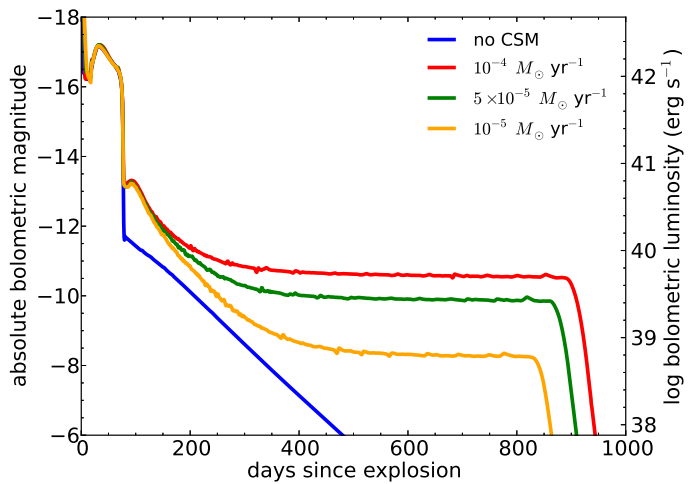}
\caption{
Bolometric LCs of ecSNe exploding within the super-AGB wind.
The LC model without CSM (TBN13) is also shown for comparison.
The mass-loss rate of the model is indicated in the figure.
The wind velocity is $v_\w =20$ $\mathrm{km~s^{-1}}$ for all models.
}
\label{bolometric}
\end{figure}

Synthetic bolometric LCs are presented in Fig. \ref{bolometric}.
The LC without CSM is the one already presented in TBN13.
It stays bright until the photosphere located at the hydrogen
recombination front resides in the H-rich envelope. This early phase is
essentially the same as the plateau phase of Type IIP SNe.
As is already shown by TBN13, the plateau phase is bright in spite of the low
explosion energy. This is because of the large progenitor radius and the
small H-rich envelope mass expected for
super-AGB SN progenitors \citep[cf.][]{kasen2009,young2004}.
The length of the plateau phase depends on the mass and composition of
the H-rich envelope and it can be longer than 100 days (TBN13).
The luminosity drops by about two orders of magnitudes
at the end of the plateau and the LC
starts to follow the decay of \Co\ if there is no CSM.

Even if we take the effect of the CSM interaction into account,
the early LCs during the plateau phase are still dominated by the
light from the SN ejecta.
The effect of the super-AGB wind on the LCs starts to appear after the plateau.

\begin{figure}
\centering
\includegraphics[width=\columnwidth]{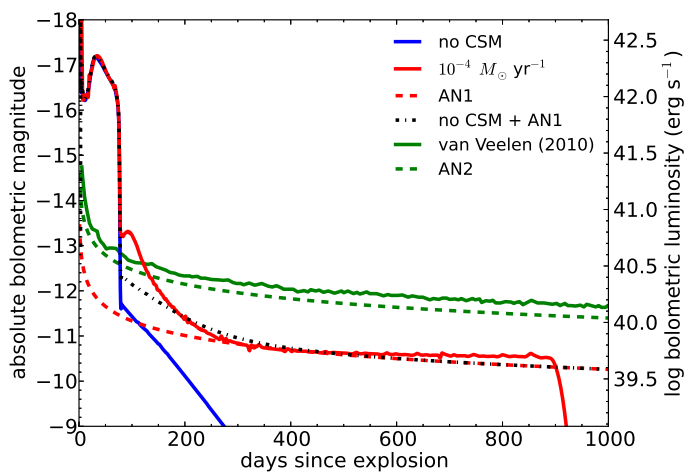}
\caption{
Comparison of several bolometric LC models.
The `no CSM' model and the $10^{-4}$ $M_\odot~\mathrm{yr^{-1}}$ model are
the same as in Fig. \ref{bolometric}. The AN1 model is the expected
 contribution to the luminosity from the interaction
in the $10^{-4}$ $M_\odot~\mathrm{yr^{-1}}$ model estimated with
the analytic LC model of \citet{moriya2013}. Adding the `no CSM' and
AN1 models, we obtain the `no CSM + AN1' model which roughly matches
the overall LC evolution of the $10^{-4}$ $M_\odot~\mathrm{yr^{-1}}$ model.
The LC model of \citet{vanveelen2010} is obtained from a numerical
two-dimensional LC calculation
 with the similar initial conditions in the $10^{-4}$
 $M_\odot~\mathrm{yr^{-1}}$ model. The AN2 model is the analytic LC estimate
 for the \citet{vanveelen2010} model. See Section \ref{secLC} for details.
}
\label{bolometriccomparison}
\end{figure}

The effect of the interaction on the LCs is examined in detail
in Fig. \ref{bolometriccomparison}.
In addition to the models with no CSM and $10^{-4}$ $M_\odot~\mathrm{yr^{-1}}$
presented in Fig. \ref{bolometric}, 
the analytic bolometric LC models based on \citet{moriya2013}
are shown (AN1 and AN2). The analytic LC is expressed as

\begin{equation}
L=\frac{\epsilon A}{2}\left(\frac{\dot{M}}{v_\w}\right)^{\frac{n-5}{n-2}}
E_\ej^{\frac{3(n-3)}{2(n-2)}}
M_\ej^{-\frac{3(n-5)}{2(n-2)}}
t^{-\frac{3}{n-2}}, \label{analytic}
\end{equation}
where
\begin{equation}
A=\left(\frac{n-3}{n-2}\right)^3
\left[
\frac{2}{(n-4)(n-3)(n-\delta)}
\frac{[2(5-\delta)(n-5)]^{(n-3)/2}}{[(3-\delta)(n-3)]^{(n-5)/2}}
\right]^{\frac{3}{n-2}},
\end{equation}
$t$ is the time since the explosion,
$E_\ej$ is the SN kinetic energy, $M_\ej$ is the SN ejecta mass,
$n$ is the outer density slope of the SN ejecta ($\rho_\ej\propto r^{-n}$),
$\delta$ is the inner density slope of the SN ejecta
($\rho_\ej\propto r^{-\delta}$),
and $\epsilon$ is the conversion efficiency from kinetic energy to radiation.
We refer \citet{moriya2013} for further details.
In this paper, we set $n=9$ and $\delta=0$ to compare the model with
that of \citet{vanveelen2010} below but the LC does not strongly depend
on the choice of $n$ and $\delta$ within the reasonable range
($n\simeq 7-12$ and $\delta\simeq 0-1$).
All the analytic models shown in Fig. \ref{bolometriccomparison} assume
$\dot{M}=10^{-4}$ $M_\odot~\mathrm{yr^{-1}}$ and $v_\w=20$ $\mathrm{km~s^{-1}}$.
The AN1 model in Fig. \ref{bolometriccomparison} assumes
$E_\ej=1.5\times 10^{50}$ erg and $M_\ej=3$ $M_\odot$.
As will be discussed below, the conversion efficiency $\epsilon$ is
uncertain in both the numerical and analytic models.
$\epsilon$ in AN1 is chosen to match the later luminosity of
the numerical $10^{-4}$ $M_\odot~\mathrm{yr^{-1}}$ model ($\epsilon=0.1$).

The analytic model represents the contribution of the
luminosity from the interaction. In Fig. \ref{bolometriccomparison},
we also present a model in which
the luminosities of the numerical no CSM model and the analytic AN1
model are added. This model can roughly explain the bolometric
luminosity evolution of the numerical model with the super-AGB wind.
Large differences between the numerical model and the added model
appear in two phases, i.e., (i) just after the plateau phase at around 100 days
and (ii) around 900 days when the numerical LC drops significantly.
The earlier luminosity discrepancy (i) is caused by the higher photospheric
temperature in the numerical interacting model.
The similar kind of smooth transition is also found in the explosion
modeling of red supergiants within dense CSM \citep{moriya2011}.
The later sudden decline (ii) appears because the shock goes out of the
dense CSM whose radius is arbitrarily set at $2\times 10^{16}$ cm.
If the super-AGB phase continues for $\sim 10^{4}$ years with the wind
speed of 20 $\mathrm{km~s^{-1}}$, the dense CSM should actually be
extended up to $\sim 6\times 10^{17}$ cm. This means that the
interaction should actually continue further and the LC continues to be
bright as we see in the analytic model.

In Fig. \ref{bolometriccomparison},
we also show the numerical bolometric LC for the interaction between
ecSN ejecta and dense CSM obtained by \citet{vanveelen2010}.
The bolometric LC is obtained with the hydrodynamics code \texttt{ZEUS-MP}
combined with a cooling function which is applicable to the optically
thin system. The method is similar to that adopted by \citet{vanmarle2010}.
The homologous SN ejecta is assumed to have two density components
with $n=9$ and $\delta=0$.
The hydrodynamical calculation is performed in two dimensions.
The SN ejecta mass is 3.15 $M_\odot$ and
the kinetic energy is $10^{50}$ erg, respectively.
These parameters are almost the same as in our model.
The SN ejecta is connected to the dense CSM with $\dot{M}=10^{-4}$
$M_\odot~\mathrm{yr^{-1}}$ and $v_\w=20$ $\kmps$ at $10^{14}$ cm.

Since the numerical interaction model of \citet{vanveelen2010} 
does not take account of the recombination process in the SN ejecta, 
the early LC evolution is different from what we obtained.
However, even at the epochs when our LC model is dominated by the
luminosity from the interaction, there remains the difference in the
luminosity by a factor of about 3.
The difference is likely to stem from the uncertainty in the
conversion efficiency $\epsilon$.
We plot another analytic LC of \citet{moriya2013} (Eq. \ref{analytic})
with $M_\ej=3.15$ \Msun\ and $E_\ej=10^{50}$ erg in
Fig. \ref{bolometriccomparison} (AN2).
The analytic model AN2 matches the numerical model of \citet{vanveelen2010}
well.
The $M_\ej$ and $E_\ej$ in AN2 are almost the same as in AN1.
The main difference between AN1 and AN2
is that the conversion efficiency $\epsilon$ is increased to 0.5 in AN2
from 0.1 in AN1.
Thus, the main reason for the discrepancy in the two numerical models
is the difference in 
the conversion efficiency achieved in these simulations.

The conversion efficiency is a parameter which is still not well-constrained.
It is usually assumed to be $\simeq 0.1-0.5$ in the literature.
The maximum conversion efficiency is $\simeq 0.5$ because of the
conservation of momentum.
In the one-dimensional code \texttt{STELLA},
the conversion efficiency is artificially reduced.
The motivation for the reduction in the code is that
the multi-dimensional instabilities may initiate
additional motion in the dense shell and
more energy may remain as kinetic energy
if the dimension is larger
(\citealt{blinnikov1998,moriya2013b}).
A recent multi-dimensional radiation hydrodynamics calculation
of a pulsational pair-instability SN model of \cite{woosley2007}
by \citet{chen2014} showed that the resulting luminosity
is similar to that obtained by \texttt{STELLA} with the artificial
reduction of the conversion efficiency.
In the multi-dimensional simulations by \citet{vanveelen2010}, however,
it is found that
the Rayleigh-Taylor fingers formed by multi-dimensional instabilities
make the density ahead of the shock higher and thus
the luminosity is higher by a factor of 1.8 in the two-dimensional
simulations.
In summary, the artificial reduction in the conversion efficiency in
\texttt{STELLA} and the larger effective conversion efficiency
in multi-dimensional simulations
by \citet{vanveelen2010} due to the Rayleigh-Taylor fingers cause
the difference in the conversion efficiency.
We still do not know clearly how the multi-dimensionality affects the
conversion efficiency.
One possibility is that the conversion efficiency depends on the CSM density.
The numerical simulations by \citet{chen2014}
are performed with the denser CSM than those in \citet{vanveelen2010}
which are closer to ours.
In this case, the conversion efficiency may also be time dependent
since the CSM density reduces as the shock propagates outward.
Another possible reason for the difference is
that the radiation is not coupled to hydrodynamics
in the simulations by
\citet{vanveelen2010} while it is coupled in the simulations
by \citet{chen2014}. The effect of the multi-dimensionality in SN IIn
luminosity needs to be investigated more.

\begin{figure}
\centering
\includegraphics[width=\columnwidth]{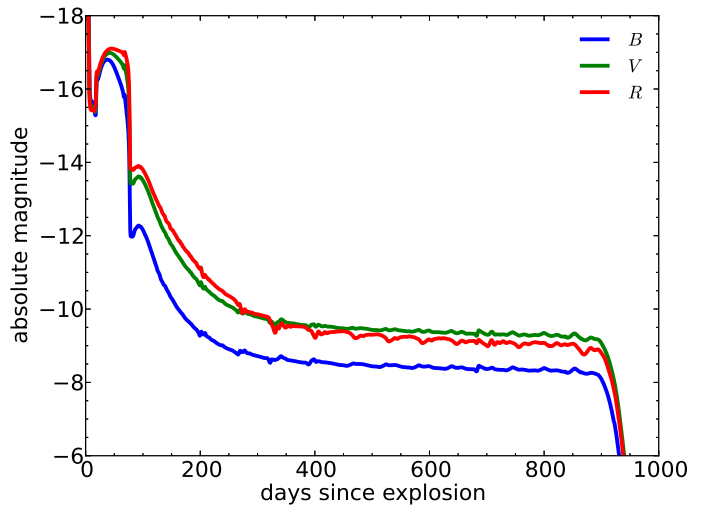}
\caption{
Multicolor LC of the model with $10^{-4}$ $M_\odot~\mathrm{yr^{-1}}$.
}
\label{multicolor}
\end{figure}

\begin{figure}
\centering
\includegraphics[width=\columnwidth]{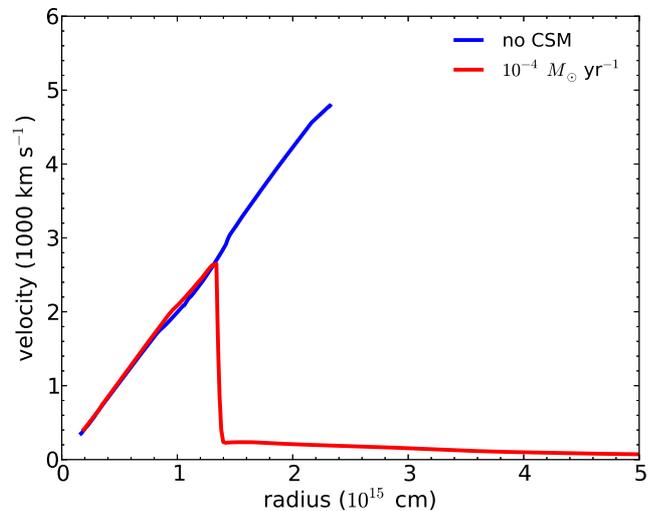}
\caption{
Velocity structures of the models with and without CSM
at $t\simeq50$ days.
}
\label{dynamics}
\end{figure}

Fig. \ref{multicolor} shows the multicolor LC of our numerical model
with $10^{-4}$ $M_\odot~\mathrm{yr^{-1}}$.
The evolution during the early plateau phase is consistent with those of
SNe IIP.
After the early plateau, the evolution of multicolor LCs is
similar to each other but the magnitudes depend on the bands.
The evolution of the optical LCs is similar to that of the bolometric LC
in Fig. \ref{bolometric}.
This is because when the shock is radiation-dominated
(when the ratio of the radiation pressure to the gas pressure is above $\simeq 5$),
the shock temperature does not become very high
because of the large heat capacity of photons and optical photons
are mainly emitted from the shock \citep{weaver1976,blinnikov2011,tolstov2013}.
Even when the shock is not radiation-dominated,
the high-energy photons produced at the
shock are efficiently thermalized due to the large CSM density and
photons can be emitted in the optical wavelengths \citep[cf.][]{chevalier2012}. 

Finally, we show the velocity structures of the models with and without
CSM in Fig. \ref{dynamics} at around
the middle of the plateau phase ($t\simeq 50$ days).
The entire velocity in the model without CSM is homologous
while only the unshocked SN ejecta is homologous in the model with CSM.
The photosphere is located at $r=1.1\times 10^{15}$ cm
in the no CSM model and it is located at $r=1.2\times 10^{15}$ cm
in the model with the CSM.
The photospheric velocity is almost the same $(2300\ \mathrm{km~s^{-1}})$
in the two models. 
The model with CSM has the velocity drop at $r=1.4\times 10^{15}$ cm
where the SN ejecta collides to the CSM. The material in SN ejecta
which exists beyond this radius in the no CSM model is accumulated
in the cold dense shell at $1.4\times 10^{15}$ cm in the model with CSM.

\section{Comparison with observations}\label{secobs}
In this section, we compare our ecSN LC models presented in the previous
section to the observations which may be related to ecSNe.

\subsection{SN 1054 (Crab Nebula)}\label{sec1054}
SN 1054 was a `guest star' which is recorded mainly in Chinese and
Japanese literature (see \citealt{stephenson2002} for a summary of
the historical records).
The proximity of the `guest star' in the historical records 
to the Crab Nebula (M 1 or NGC 1952) was first pointed out by
\citet{lundmark1921}. The expansion of the Crab Nebula
which was discovered by \citet{lampland1921,duncan1921}
is consistent with the explosion occurring in 1054
(\citealt{hubble1928,rudie2008}, but see also
\citealt{trimble1968,wyckoff1977,nugent1998}).
Thus, the Crab Nebula is quite likely the remnant of SN 1054.

The Crab Nebula is one of the best observed SN remnants
(see \citealt{hester2008,davidson1985} for summaries).
The observed chemical abundances
\citep[e.g.,][]{macalpine2008,satterfield2012}
suggest that the Crab Nebula is
a remnant of an ecSN \citep[e.g.,][]{nomoto1982}.
In addition, the estimated small ejecta mass ($4.6\pm1.8\ M_\odot$,
\citealt{fesen1997})
and small kinetic energy
($\sim 10^{49}$ erg, e.g., \citealt{frail1995,smith2013})
match the theoretical predictions of ecSN properties
\citep[e.g.,][]{nomoto1982,kitaura2006,burrows2007}.
The above properties of the Crab Nebula suggest that SN 1054 was likely an ecSN.

SN LCs are an important clue to investigate their progenitors.
The LC of SN 1054 deduced from the ancient records (see below)
has been extensively discussed by many
authors \citep[e.g.,][]{minkowski1971,chevalier1977,wheeler1978}.
It has been argued that the LC of SN 1054 may be
inconsistent with those expected from ecSNe.
\citet{sollerman2001} pointed out that the last
observational record of SN 1054 is incompatible with the small amount of
\Ni\ production expected in ecSNe.
Recently, \citet{smith2013} suggested that the relatively high early
luminosity of SN 1054 is hard to be explained by the small explosion energy
expected for ecSNe. However, TBN13 showed that the small explosion
energy is consistent with the early high luminosity of SN 1054
because of the progenitor's large radius and the small H-rich envelope
mass in ecSNe.
\citet{sollerman2001} proposed that the later high luminosity of SN 1054
was powered by
energy input from the central pulsar or the CSM interaction.
TBN13 took the energy input from the pulsar into account. They found
that the energy input from the pulsar can potentially explain the late
luminosity, although their model with one-group radiation transport predicted
a lower luminosity.
We here compare our ecSN LC models with super-AGB winds with that of
SN 1054 to show that ecSN exploded within super-AGB winds can actually explain
the LC properties of SN 1054.

\begin{figure}
\centering
\includegraphics[width=\columnwidth]{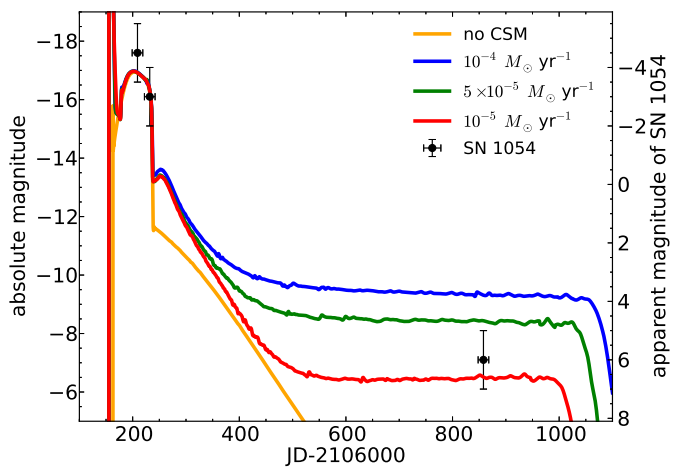}
\caption{
LCs of SN 1054 and our ecSNe exploded within super-AGB winds.
The $V$ band LCs of the models are shown. The observations of SN 1054
are based on human eyes which are sensitive at around the $V$ band
\citep{vos1978}.
Note that the $V$ band and $R$ band
magnitudes are similar in our models (Fig. \ref{multicolor}).
}
\label{sn1054}
\end{figure}

In Fig. \ref{sn1054}, we compare our LC models with the LC of SN 1054
deduced from the historical records. 
The recorded brightness of SN 1054 at three epochs which were observed with human eyes
are adopted \citep{stephenson2002}:
\begin{enumerate}
\item The `guest star' appeared on AD 1054 July 4 (JD 2106219)\footnote{
The star was recorded to appear in late May in 1054 in Japanese literature
(\textit{Meigetsuki}) but it is likely that there is an error in the
recorded month and the actual date is one month later which is consistent
 with the Chinese records \citep{stephenson2002}.}
and it was as bright as Venus ($\sim -4.5$ mag).
\item It was able to observe during the daytime for 23 days.
Thus, the star was as bright as the sky in the daytime
at 23 days since the discovery (JD 2106232), i.e.,
$\sim -3$ mag.
\item It disappeared from the sky on AD 1056 April 6 (JD 2106858).
We assume that SN 1054 became $\sim 6$ mag at this time.
\end{enumerate}
The errors, especially in the magnitudes, are expected to be large.
For example, the Japanese record reports that the star was as bright as
Jupiter ($m_\mathrm{opt}\sim -2$ mag), rather than Venus \citep{stephenson2002}.
We assign one magnitude error to the deduced magnitudes.
The recorded date also has uncertainty. In addition, the star may not
have been seen for a while because of the bad weather.
Thus, we assign the error of 10 days in the recorded dates.
The estimated observed magnitudes are corrected to the absolute ones with
the distance of 2 kpc \citep{trimble1973,davidson1985}
and
the extinction of $A_V=1.6$ mag \citep{miller1973}.

Our LC models show that the explosion of an ecSN within super-AGB wind
can explain the major characteristics of SN 1054.
As has already been shown by TBN13, the early high luminosity of SN 1054
can be explained by the ecSN model.
\citet{smith2013} suggested that a confined dense shell is required
to explain the early high luminosity of SN 1054 with a low energy
explosion of ecSNe but this is not necessarily the case (TBN13).
The early luminosity can be solely explained by ecSNe without any
confined dense shell.
No extreme mass loss of the super-AGB star shortly before the
explosion is required.

An essential difference between our models and those of TBN13 is the
existence of the super-AGB wind in our models. As was discussed by
\citet{sollerman2001}, the late luminosity of SN 1054 requires
about 0.06 \Msun\ of \Ni, which is incompatible 
with the expected small \Ni\ production of ecSNe ($\sim 10^{-3}$ \Msun). 
TBN13 invoked the luminosity input from the central pulsar to explain
the late high luminosity. 
Our models including the luminosity from the interaction with dense CSM
whose existence is naturally expected in the explosion of super-AGB
stars 
can explain the later luminosity of SN 1054 without invoking the
energy input from the pulsar.
This possibility was proposed already by \citet{sollerman2001} and
we confirmed it.
Both the pulsar and interaction may contribute to the late time luminosity.

The difference in observational properties of the pulsar and interaction
models appears in the spectra. The interaction is presumed to make SN
1054 Type IIn. The luminosity from the interaction dominates after the
early plateau phase but the strong narrow emission lines 
may have made SN 1054 Type IIn from early on. 
Observations of light echoes may be able to distinguish them as was
suggested by \citet{smith2013}. However,
detection of SN IIn spectra does not necessarily indicate that
SN 1054 was powered by the interaction from the earliest time as is
shown here.

A remaining concern about our LC models for SN 1054 is the final
observational record. The final LC point is based on the Chinese record
that the `guest star' vanished at that epoch. 
However, the SN luminosity of our models stays constant.
One can, of course, doubt the accuracy of the record and the luminosity
may have stayed more or less similar for longer time.
Another possible interpretation is that the shock had gone through the
dense part of the CSM at this time and the LC declined suddenly as
seen at around JD 2107000 in Fig. \ref{sn1054} in our models whose
dense CSM radius is $2\times 10^{16}$ cm. The disappearance at JD 2106858
can be explained if the dense CSM only extends to
$\simeq 1.5\times 10^{16}$ cm. In this case, the super-AGB progenitor should
have undergone the high mass-loss rate starting from $\simeq$ 240 years before
the explosion, assuming $v_\w=20\ \kmps$.
This may be possible if the mass loss from super-AGB stars
is related to many thermal pulses of He shell burning, thus
being strongly time-dependent. 

\begin{figure}
\centering
\includegraphics[width=\columnwidth]{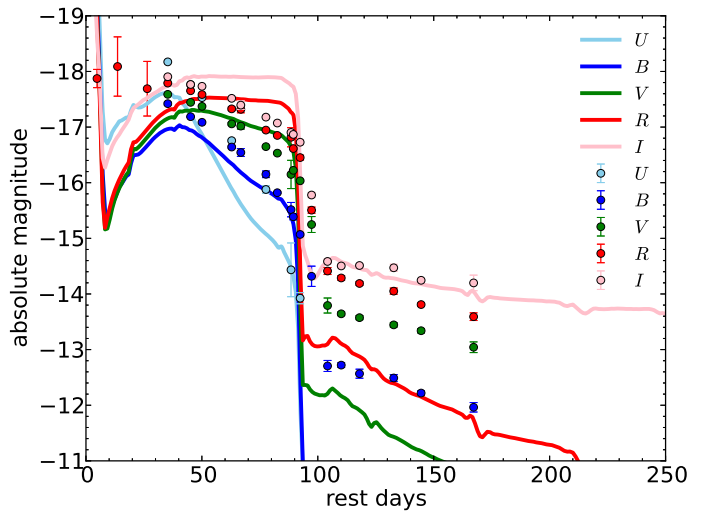}
\caption{
Comparison of the observed LC of SN 2009kn \citep{kankare2012} and
the multicolor LC model of the ecSN exploded in the super-AGB wind.
}
\label{sn2009kn}
\end{figure}

\subsection{Type IIn supernovae with LC plateau (`Type IIn-P')}
Some SNe IIn are known to be characterized by an early LC plateau
followed by the sudden LC decline as seen in SNe IIP.
They have been suggested to be
related to ecSNe because of the low estimated \Ni\ mass
(e.g., SN 1994W, \citealt{sollerman1998};
SN 2009kn, \citealt{kankare2012}; SN 2011ht, \citealt{mauerhan2013}).
This kind of SNe IIn is also called `Type IIn-P' SNe
by some authors because of the LC plateau \citep[e.g.,][]{mauerhan2013}.

In Fig. \ref{sn2009kn}, we compare the ecSN LC model interacting with
a super-AGB wind with the LC of SN 2009kn \citep{kankare2012}.
The progenitor model is constructed by attaching the
$\dot{M}=10^{-4}$ $M_\odot~\mathrm{yr^{-1}}$ and $v_\w =20$ $\mathrm{km~s^{-1}}$
wind on top of the $M_\mathrm{env}=4.7\ M_\odot$ and $X_\mathrm{env}=0.5$
progenitor presented in TBN13. 
The explosion energy is increased to $3.5\times 10^{50}$ erg
to match the plateau luminosity of SN 2009kn. The \Ni\ mass is the
same as in the previous models ($2.5\times 10^{-3}$ $M_\odot$) for simplicity.

The overall LC properties of SN 2009kn are reproduced by the numerical
model. The $R$-band brightness of the first three observed epochs is
higher than the numerical LC. However, the multicolor LCs of the
later plateau phase are 
consistent with the observations. The luminosities
after the plateau are lower than observed ones.
This may indicate that the mass-loss rate of the progenitor was larger
than $10^{-4}$ $M_\odot~\mathrm{yr^{-1}}$ and/or the wind velocity
was smaller than $20$ $\mathrm{km~s^{-1}}$.

For the LC drop after the plateau, two reasons have been suggested.
(i) \citet{chugai2004} suggested that the drop appears when the blast wave
leaves the dense part of CSM.
The major power input from the interaction terminates when the blast
wave leaves the dense part of the CSM and the LC drop is observed.
(ii) \citet{dessart2009} related the drop to the recombination.
Our model is close to the latter idea.
In our model, however, 
the LC plateau is not related to the interaction as considered in
\citet{dessart2009}, but caused by exactly the same reason as in
SNe IIP, i.e., the recombination in the H-rich layers of the SN ejecta.

As the emission from the early phase is dominated by the emission from
the SN ejecta, one may expect to see SN IIP spectral features in these
epochs.
However, the photospheric velocity during the plateau phase is as low
as $\sim 2000$ $\mathrm{km~s^{-1}}$
(Section \ref{secLC} and Fig. \ref{dynamics}).
Thus, the spectral features from the SN ejecta can be strongly diluted
by the intermediate narrow-line components usually observed in SNe IIn
which typically have the width of $\sim 1000$ $\mathrm{km~s^{-1}}$.
The dense shell may also smear out the spectral
features from the SN ejecta. 

One observational property of these SNe IIn which may not prefer our
ecSN model is the observed velocities of the
narrow P-Cygni profiles. The small velocities of the narrow P-Cygni
profiles in SNe IIn are generally related to the wind velocity of the
progenitors. The velocities of the narrow components in
the SNe IIn with the LC plateau are
typically $100-1000$ $\kmps$
\citep[e.g.,][]{chugai2004,kankare2012,mauerhan2013}.
This is inconsistent with the small wind velocities of the super-AGB
wind. One possibility is that those features may come from the
slowly-moving SN ejecta and there exist other unobserved much slower P-Cygni
components in the spectra.
In addition, the detailed line
formation mechanisms of SNe IIn are not well-understood and the detailed
spectral modeling is required to see if the super-AGB model is able to explain
the spectral properties.

Finally, one of the SNe IIn with the LC plateau, SN 2011ht, is
found to have brightened a year before the major luminosity increase
in 2011 \citep{fraser2013}. If SN 2011ht is associated with a super-AGB
star, the pre-outburst could be related to the thermal pulses of
He-shell burning in
the super-AGB progenitor. Several possible pre-burst events have been reported
\citep{ofek2014} and some of them may be related to the AGB activities.

\subsection{Other supernovae}
Other SNe IIn which have been suggested to be ecSNe are faint from early times.
For example, the peak bolometric magnitude of SN 2008S was about $-14$
mag ($10^{41}$ erg, e.g., \citealt{botticella2009})
and the progenitor was found to be around 10 $M_\odot$ \citep{prieto2008}.
Our LC models, however, predict that ecSNe are not faint at early times.
One possible way to make the early ecSN LCs faint is to reduce
the mass of H-rich envelope.
Then, the early bright phase gets as short as those in SNe IIb and
the subsequent luminosity is dominated by the interaction from early times
because of the small $^{56}$Ni mass. The luminosities only from the
interaction in our models
are comparable to those of faint SNe IIn which have been
suggested to be ecSNe like SN 2008S
(Fig. \ref{bolometriccomparison}). The H-rich envelope may become small enough
depending on the amount of mass loss from the progenitors
if they are single stars. ecSNe from binary stars can also lose large
amount of mass due to the binary evolution \citep[cf.][]{podsiadlowski2004}.
Without the super-AGB wind,
faint SNe IIn, such as SN 2008S, may have been observed as SNe IIb
from ecSNe with very small mass H-rich envelopes.
In some cases, the SN IIb features may be observed at early times and
ecSNe with little H-rich envelope may be observed as `Type IIn-b'.

X-ray observations of SNe IIn can play an important role to identify
the SNe IIn of the ecSN origin. Although our ecSN LCs with the plateau
phase are optically bright
at early times, the high luminosity is not due to the interaction.
This means that X-ray luminosities which are mainly from the interaction
can be very low compared to other SNe IIn with similar optical luminosities
which are powered mainly by the interaction in the optical range as well.
For example, the X-ray luminosities of 
$\sim 10^{37}$ $\mathrm{erg~s^{-1}}$ in $0.1-10$ keV at one year since
the explosion is expected in our $10^{-4}$ $M_\odot~\mathrm{yr^{-1}}$
model with the formalism of \citet{fransson1996}.
SN IIn 2005ip, which had a similar optical luminosity to our ecSN model
at the early plateau \citep{stritzinger2012},
had the X-ray luminosity of $\simeq 10^{41}$ $\mathrm{erg~s^{-1}}$
in $0.1-10$ keV at one year after the explosion \citep{katsuda2014}.
Thus, optically-bright but X-ray-faint
SNe IIn can be promising candidates for ecSNe.
However, if SN ejecta mass is small due to the mass loss or
binary evolution, the luminosity from the interaction can be high
(\citealt{vanveelen2010}, see Eq. \ref{analytic}).
In other words, the luminosities of ecSNe with little H-rich envelope would
be dominated by the interaction from early times and
their X-ray luminosities can be as bright as those expected from
the optical luminosities.
Then, the corresponding SNe IIn can also be X-ray bright for a long time
as is observed in, e.g., SN 1988Z \citep{vanveelen2010}.
Radio observations may also be used to trace the shock properties but
the radio luminosity can be low because of the strong free-free
absorption by the dense CSM \citep[cf.][]{chevalier1998}.

\section{Conclusions}\label{secconclusion}
We performed numerical LC calculations of ecSNe exploding within
super-AGB winds and obtained the following results.
The early bolometric LCs at the plateau phase are not affected by the dense CSM
and they are the same as those from the models without the dense CSM (TBN13).
As is shown by TBN13, the early LCs of ecSNe can be as luminous as
$-17$ mag in the optical bands in spite of the small explosion energy
($1.5\times 10^{50}$ erg) because of the large radius
and the small H-rich envelope mass of super-AGB stars.
The effect of the super-AGB CSM on the LCs starts to appear after the plateau
phase. If there is no CSM, the LCs start to follow the $^{56}$Co decay
($L\propto e^{-t/(111\ \mathrm{days})}$) after the plateau.
However, if there exists a dense CSM from the super-AGB stars,
the interaction between the SN ejecta and the dense super-AGB CSM
provides additional energy to illuminate ecSNe.
The bolometric LCs decline much slower than the \Co\ decay
($L\propto t^{-3/(n-2)}$,
where $n$ is the outer density slope of SN ejecta,
$\rho_\ej\propto r^{-n}$, with $n\simeq 7-12$).

SN 1054 (Crab Nebula) has been suggested to be an ecSN.
However, several observational properties have been argued to be difficult
to be explained by ecSNe.
\citet{sollerman2001} argued that the final observational record
at $t\simeq650$ days after the discovery
indicates that SN 1054 was too bright to be an ecSN which does not produce
enough $^{56}$Ni.
We confirmed that the interaction with the dense CSM expected from the
super-AGB stars can explain the last observational record.
\citet{smith2013} argued that the early luminosity is also too bright
to be ecSNe. However, it was shown by TBN13 as well as in this work
that the luminosity is consistent with ecSNe.

We also compare our LC model to the LCs of
a subclass of SNe IIn which show the LC plateau in the early phase
(`Type IIn-P').
We show that our LC model is consistent with them and suggest that
they are ecSNe exploding within the dense super-AGB CSM.
Faint SNe IIn can be ecSNe with the dense CSM but
with little H-rich envelope. They may be observed as `Type IIn-b'.

We show that SNe IIn with the LC plateau from ecSNe
can be bright in optical at early times
even if the CSM is not dense enough to power the early optical luminosity
by the interaction.
Thus, their X-ray luminosities from the interaction can be
faint in spite of the large optical luminosities in the SNe IIn.
As a result, some ecSNe can be
observed as optically-bright but X-ray-faint SNe IIn.
Since the luminosities of ecSNe with little H-rich envelope
are dominated by the interaction from early on,
their X-ray luminosities can be as bright as those expected from
the optical luminosities.

\begin{acknowledgements}
TJM is supported by Japan Society for the Promotion of
 Science Postdoctoral Fellowships for Research Abroad
 (26\textperiodcentered 51).
KN is supported by the Grant-in-Aid for Scientific Research (23224004, 23540262, 26400222) from the Japan Society for the Promotion of Science.
Numerical computations were carried out on computers at Center for
 Computational Astrophysics, National Astronomical Observatory of Japan.
This research is also supported by World Premier International Research Center Initiative (WPI Initiative), MEXT, Japan.
The work in Russia was supported by RF Government grant 11.G34.31.0047, grants for supporting Scientific Schools 5440.2012.2 and 3205.2012.2, and joint RFBR-JSPS grant 13--02--92119.
\end{acknowledgements}

\end{document}